\documentclass[]{aastex}
\usepackage{psfig,amsfonts,amsmath,graphicx,natbib,apjfonts,lscape,nicefrac,
url,hyperref,longtable,threeparttable}

\usepackage{siunitx,booktabs}
\hypersetup{colorlinks=true, urlcolor=blue, citecolor=blue}

\newcommand{\Swift}{\textit{Swift}}
\newcommand{\Fermi}{\textit{Fermi}}

\newcommand{\numax}{$\nu_{\rm m}$}

\newcommand{\epse}{$\epsilon_{\rm e}$}
\newcommand{\epsb}{$\epsilon_{\rm b}$}

\newcommand{\emcee}{\textsc{emcee}}
\newcommand{\me}{GRB~130427A}

\shorttitle{GRB~130427AA}

\def\cfa{1}
\def\tifr{2}

\begin{document}
\title{A Reverse Shock in GRB~130427A}

\author{
  T. Laskar\altaffilmark{\cfa},
  E. Berger\altaffilmark{\cfa},
  B. A. Zauderer\altaffilmark{\cfa},
  R. Margutti\altaffilmark{\cfa},
  A.~M. Soderberg\altaffilmark{\cfa},
  S. Chakraborti\altaffilmark{\cfa},
  R. Lunnan\altaffilmark{\cfa},
  R. Chornock\altaffilmark{\cfa},
  P. Chandra\altaffilmark{\tifr},
  and A. Ray\altaffilmark{\tifr}
}

\altaffiltext{\cfa}{Harvard-Smithsonian Center for Astrophysics, 60
Garden Street, Cambridge, MA 02138}

\altaffiltext{\tifr}{Tata Institute of Fundamental Research,
Homi Bhabha Road, Mumbai 400 005, India}

\shortauthors{Laskar et al.}

\begin{abstract}
We present extensive radio and millimeter observations of the
unusually bright GRB\,130427A at $z=0.340$, spanning 0.67 to 12~d
after the burst.  Taken in conjunction with detailed multi-band UV,
optical, NIR, and X-ray observations we find that the broad-band
afterglow emission is composed of distinct reverse shock and forward
shock contributions.  The reverse shock emission dominates in the
radio/millimeter and at $\lesssim 0.1$ d in the UV/optical/NIR, while
the forward shock emission dominates in the X-rays and at $\gtrsim
0.1$ d in the UV/optical/NIR.  We further find that the optical and
X-ray data require a Wind circumburst environment, pointing to a
massive star progenitor.  Using the combined forward and reverse shock
emission we find that the parameters of the burst are an isotropic
kinetic energy of $E_{\rm K,iso}\approx 2\times 10^{53}$ erg, a mass loss
rate of $\dot{M}\approx 3\times 10^{-8}$ M$_\odot$ yr$^{-1}$ (for a
wind velocity of $1,000$ km s$^{-1}$), and a Lorentz factor at the
deceleration time of $\Gamma(200\,{\rm s})\approx 130$.  Due to the
low density and large isotropic energy, the absence of a jet break to
$\approx 15$ d places only a weak constraint on the opening
angle, $\theta_j\gtrsim 2.5^\circ$,
and therefore a total energy of $E_{\rm \gamma}+E_K\gtrsim
1.2\times 10^{51}$ erg, similar to other GRBs.  The reverse shock
emission is detectable in this burst due to the low circumburst
density, which leads to a slow cooling shock.  We speculate that this
is a required property for the detectability of reverse shocks in the
radio and millimeter bands.  Following on GRB\,130427A as a benchmark
event, observations of future GRBs with the exquisite sensitivity of
VLA and ALMA, coupled with detailed modeling of the reverse and
forward shock contributions will test this hypothesis.
\end{abstract}

\section{Introduction}
Although long-duration $\gamma$-ray bursts (GRBs) have been associated with the deaths of massive
stars \citep{wb06}, the precise nature of their progenitors, the structure of their explosion
environments, and the composition of their ejecta remain only partially explored.   Studies of the
afterglow (forward shock) emission provide insight on the explosion energy, geometry, and the
structure of the circumburst medium (e.g., \citealt{spn98,cl00}).  On the other hand,
the most useful probe of the initial bulk Lorentz factor and the ejecta composition is afforded by
the reverse shock, which is expected to produce a similar synchrotron spectrum as the forward shock,
with well-defined properties relative to the forward shock (e.g., \citealt{sp99,kz03a,zwd05}).

The expected observational signature of the reverse shock is early time flares in the optical and
radio bands, and several studies have found hints of excess early-time emission attributable to a
reverse-shock like component \citep[e.g.][]{abb+99,sp99a,kfs+99,sr02,kz03,bsfk03,sr03,clf04}.
However, a detailed understanding of the reverse shock emission requires a careful decomposition of
the afterglow spectral energy distribution into the reverse and forward shock components.  Since the
peak frequencies of the two components are related by a factor of $\Gamma^2\gtrsim 10^4$, such a
decomposition requires multi-wavelength observations spanning several orders of magnitude in
frequency.  Similarly, the density profile of the environment affects the hydrodynamic evolution of
both the forward and reverse shocks, leading to discernible differences in the behavior of light
curves.  Consequently, well-sampled light curves are also essential for any systematic study of
reverse shocks.

Such data sets have not been available to date primarily due to sensitivity limitations of radio and
millimeter facilities. However, the recent upgrade of the VLA to the Karl G. Jansky Very Large Array
(JVLA), with an order of magnitude improvement in sensitivity has opened a new avenue for the study
of reverse shocks in GRBs. Here we present the first example of a reverse shock detected in a
multi-wavelength data set spanning 1--100 GHz of the nearby energetic GRB~130427A.  By combining our
detailed radio and millimeter observations  with X-ray data from \Swift\ and UV/optical/NIR
observations from \Swift\ and ground-based telescopes, we present the most comprehensive data set
for reverse shock studies to date.  We undertake a joint model fit to the entire data set and
extract parameters for the explosion and the environment, and draw general conclusions on reverse
shock studies in the JVLA and ALMA era.

\section{Observations and Data Analysis}
\label{sec:obs}
\subsection{GRB Discovery}
GRB~130427A was discovered by the \Swift\ \citep{gcg+04} Burst Alert Telescope
\citep[BAT,][]{bbc+05} on 2013 April 27 at 07:17:57\,UT \citep{gcn14448} with a duration of
$T_{90} = 163$\,s and a fluence of $F_{\gamma} = 3.1 \times10^{-4}$\,erg\,cm$^{-2}$
\citep[$15$--$150$\,keV;][]{gcn14470}.
The burst was also detected with the \Fermi\ Gamma-ray Burst Monitor \cite[GBM;][]{mlb+09}, 50.6\,s
before the \Swift\ trigger\footnote{In our analysis, we take the GBM trigger time as the start
time of the event, $t_0$} with an unusually large fluence of $F_{\gamma} = 2
\times 10^{-3}$\,erg\,cm$^{-2}$ (10--1000~keV) and a peak energy of $E_{\rm p} = 830$\,keV
\citep{gcn14473}. Coincident high-energy $\gamma$-ray emission was detected by the \Fermi\ Large
Area Telescope \cite[LAT;][]{aaa+09} up to $94$~GeV \citep{gcn14471}.

The \Swift\ X-ray Telescope \citep[XRT,][]{bhn+05} began observing the field 140\,s after
the BAT trigger, leading to the detection of the X-ray afterglow, localized to RA(J2000)  =
11\,$^{\rm h}$ 32\,$^{\rm m}$ 32.63\,$^{\rm s}$, Dec(J2000) = +27$^{\circ}$ 41\,' 51.7\,", with an
uncertainty radius of 3.5 arcseconds \citep[90\%
containment,][]{gcn14485}.	
The \Swift\ UV/Optical Telescope \citep[UVOT,][]{rkm+05} began observing the field 150~s
after the burst, leading to the detection of a bright UV/optical afterglow \citep{gcn14448}, which
was subsequently detected by several ground-based observatories in the optical
\citep[e.g.][]{gcn14450,gcn14451,gcn14452,gcn14453,gcn14454}, NIR \citep{gcn14453}, millimeter
\citep{gcn14482,gcn14494}, and radio \citep{gcn14480,gcn14505,gcn14522}.
Spectra were obtained at $Gemini$-North, the \textit{Nordic Optical Telescope}, and
the \textit{VLT} resulting in a redshift  of $z = 0.340$ \citep{gcn14455,gcn14478,gcn14491}, and
leading to an isotropic equivalent $\gamma$-ray energy of $(1.05\pm0.15)\times10^{54}$~erg
\citep[$E_{\gamma,{\rm iso}} = $ 1--$10^4$~keV, rest frame;][]{gcn14503}.

\subsection{Radio to X-ray Observations}
We observed the position of GRB\,130427A beginning on 2013 April 27.99 UT
($\Delta t$ = 0.67 d) with the Karl G. Jansky Very Large Array (VLA) at
a mean frequency of 5.8, GHz and with the Combined Array for Research
in Millimeter Astronomy \citep[CARMA;][]{bbh+06} beginning on 2013 April
28.13 UT ($\Delta t = 0.81$ d) at a mean frequency of 85 GHz.
In both observations, we detect a strong
radio source coincident with the optical afterglow position.  The data
were obtained in the standard continuum modes utilizing the VLA's WIDAR
correlator \citep{pcb+11} with a total bandwidth of $\sim$2 GHz, and
CARMA's continuum mode with a bandwidth of $\sim$8 GHz.  For VLA
observations we utilized 3C286 for bandpass and flux calibration, and
J1125$+$2610 for gain calibration in all but one epoch, where
J1159$+$2914 was utilized as the gain calibrator for observing
frequencies greater than 15 GHz.  For all CARMA observations we
utilized 3C273 for primary flux calibration and 0854$+$201 for
bandpass calibration.  For gain calibration, we utilized 1224$+$213,
and a source-gain cycle of $\sim$15 min.  We analyzed the VLA
observations using the Common Astronomy Software Applications
\citep[CASA;][]{mws+07} and the Astronomical Image Processing System
\citep[AIPS;][]{gre03}, and the CARMA observations using the
Multi-channel Image Reconstruction, Image Analysis, and
Display software \citep[MIRIAD;][]{stw95}.  In all cases, we
flagged edge channels and any data corrupted with radio frequency
interference.  We measured the flux density of the afterglow in the
final images using the AIPS task JMFIT.
We also observed GRB 130427A with the Giant Metrewave Radio Telescope (GMRT) at central frequencies
of 1390 MHz and 610 MHz (bandwidth of 32 MHz), using 3C\,147 and 3C\,286 as flux and bandpass
calibrators, and J1125+261 and J1227+365 as phase calibrators. We analyzed the data
using AIPS. A summary of all radio and millimeter observations is provided in Table
\ref{tab:photometry}.

We extracted \Swift/XRT spectra at the times of our radio observations (at 58, 173, 406, and
786~ks) using the latest version of the HEASOFT package (v. 6.13) and corresponding calibration
files, following standard procedures \citep{ebp+07,ebp+09,mzb+13}. The spectra are well
fit by an absorbed single power-law model with $N_{\rm H,Gal}=1.80\times 10^{20}~\rm{cm^{-2}}$
\citep{kbh+05}, intrinsic hydrogen
column $N_{\rm H,int} = (6.64\pm0.01)\times 10^{20}~\rm{cm^{-2}}$ and photon index
$\Gamma = 1.76\pm0.03$ ($68\%$ confidence intervals). We find no statistically
significant evidence for spectral evolution. We obtain XRT light curves by converting
the $0.3$--$10$\,keV count rates reported on the \Swift\
website\footnote{\url{http://www.swift.ac.uk/xrt_curves/00554620/}} to a flux density at 1\,keV
using the measured spectral model \citep{ebp+07,ebp+09}.

\Swift/UVOT observed \me\ using 6 filters spanning the central wavelength range $\lambda_{\rm
c}=1928$\,\AA\ (\textit{w2}) to $\lambda_{\rm c}=5468$\,\AA (\textit{v}).
We analyzed the UVOT data using the latest version of HEASOFT (v. 6.13) and corresponding
calibration files with a $5''$ aperture extract on in the UVOT photometric system \citep{blh+11}.
The \Swift\ star trackers failed to find a correct aspect solution during the first $1.9$~ks of
exposure. For these data, we manually checked each frame and corrected the position of the
extraction region to account for the drifting of the source . The source point spread function
appears to be highly distorted in the first two frames acquired in $v$-band, while the first frames
in the $b$ and $u$ bands suffer from significant coincidence losses due to the brightness of the
afterglow. We do not use these exposures in our analysis.

We also obtained three epochs of $g^{\prime}r^{\prime}i^{\prime}$ photometry of the optical
afterglow using the MMTcam imager on the MMT
6.5m telescope\footnote{http://www.cfa.harvard.edu/mmti/wfs.html}.
A sequence of $2
\times 150$\,s dithered images in each band was taken on 2013 May 2.43 UT, of $2 \times 250$\,s (in
$g^{\prime}$ and $r^{\prime}$) and $2\times 300$\,s (in $i^{\prime}$) on 2013 May 4.43 UT, and $3
\times 300$s in each band on 2013 May 8.44 UT. The images were bias-and dark-subtracted,
flat-fielded and stacked using standard routines in IRAF\footnote{IRAF is
distributed by the National Optical Astronomy Observatories, which are operated by the Association
of Universities for Research in Astronomy, Inc., under cooperative agreement with the National
Science Foundation.}. The combined flux of the GRB and host galaxy was measured using aperture
photometry with an aperture size of $2.5$ times the seeing of the images, and calibrated relative to
the SDSS photometry of the nearby star SDSSJ113231.32+274222.7.

Finally, we collected all publicly-available photometry from the GCN Circulars and converted
the reported magnitudes to flux densities at the following central wavelengths:
$g^{\prime}:445\,$nm, $V:551$\,nm, $R:658$\,nm, $I:806$\,nm, $z^{\prime}:900$\,nm, $Y:1020$\,nm,
$J:1220$\,nm, $H:1630$\,nm, and $K:2190$\,nm. For magnitudes reported in the Vega system, we used
the zero-points from \cite{bcp98}.

\section{Basic Considerations}
\label{sec:basic}

We interpret the observed behavior of the afterglow emission from
radio to X-rays in the framework of the standard synchrotron model, described
by three break frequencies (self-absorption: $\nu_a$; characteristic
synchrotron frequency: $\nu_m$; and synchrotron cooling: $\nu_c$),
and an overall flux normalization.
In this model, there are well-defined
relations between the temporal evolution and spectral indices that
allow us to determine the location of the synchrotron frequencies as
well as the density profile of the circumburst medium (ISM profile:
$\rho={\rm const}$; Wind: $\rho\propto r^{-2}$).  The models are
described in detail in \citet{spn98} and \citet{cl00}.

\begin{figure}[ht]
\centering
\includegraphics[width=\columnwidth]{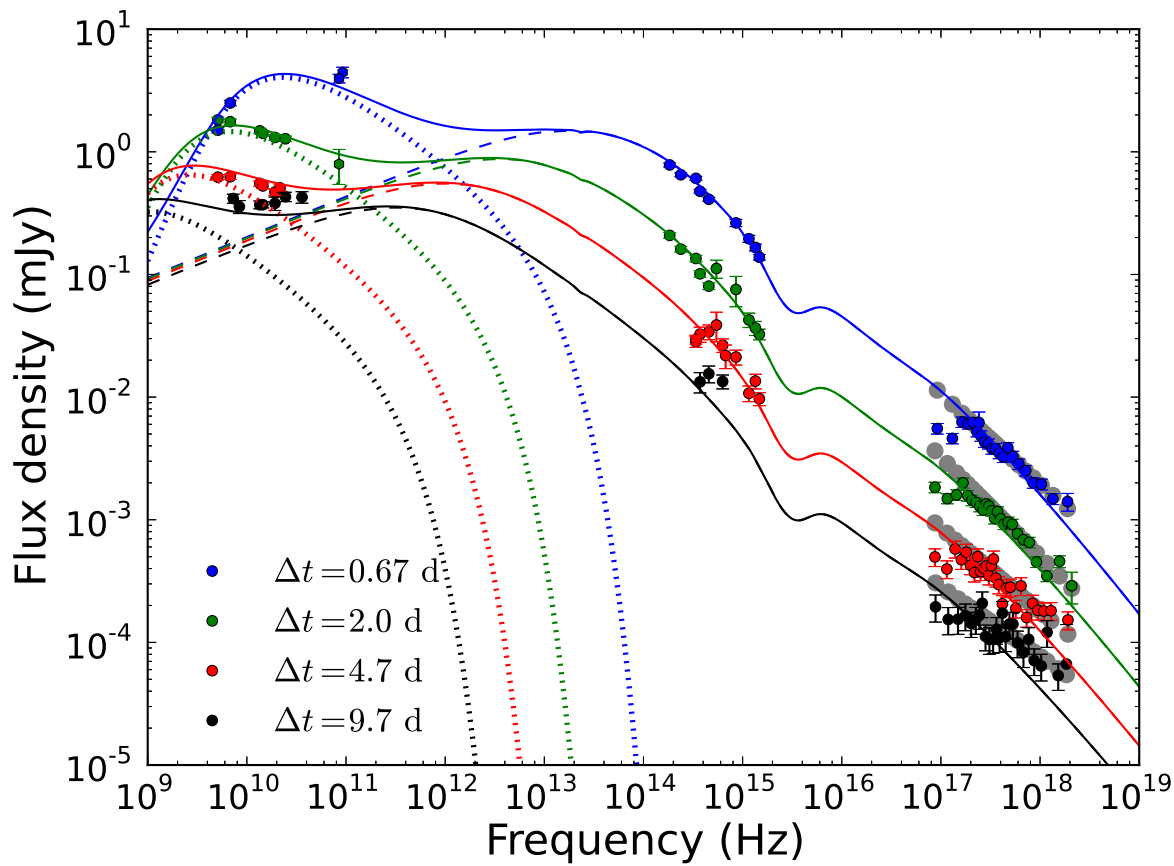}
\caption{The spectral energy distribution of the afterglow at $\Delta t = 0.67$, 2.0, 4.7, and 9.7
days. The optical data are small extrapolations from the nearest available data points with a power
law of $t^{-1.35}$. The dip in the model around $3\times10^{15}$~Hz is caused by extinction
($A_{\rm V}=0.18$~mag) in the host galaxy. The light grey points represent the unabsorbed models
for the X-ray spectra. The dashed and dotted curves show the spectrum of the forward shock and
reverse shocks, respectively, while the solid lines are the sum of the two. The combined model fully
captures the observed evolution across nine orders of magnitude in frequency.
\label{fig:sed}}
\end{figure}

A striking feature of the afterglow light curves is the unbroken
power-law decline in the X-rays with $\alpha_X\approx -1.35$ at
$\gtrsim 200$ s ($F_\nu\propto t^\alpha$).  Given the X-ray spectral
index of $\beta_X\approx -0.76$ during this time ($F_\nu\propto
\nu^\beta$), we expect an X-ray decline rate of $\alpha_X=-0.64$ if
$\nu_c<\nu_X$ (independent of the circumburst density profile), while
for $\nu_c>\nu_X$, we expect $\alpha_{\rm X}\approx -1.14$ for an ISM
profile or $\alpha_{\rm X}\approx -1.64$ for a Wind profile.  Since
the decline rate predicted in the ISM model is always shallower than
the observed value, we conclude that a Wind profile is required,
with $\nu_c\sim \nu_X$ providing a reasonable match to the observed
decline.  This conclusion is further supported by the similar decline
rate in the UV/optical/NIR band at $\gtrsim 0.3$ d, $\alpha_{\rm opt}
\approx -1.35$, which indicates that the optical and X-ray bands are
located on the same portion of the synchrotron spectrum (i.e.,
$\nu_m<\nu_{\rm opt}<\nu_X\sim \nu_c$).  Indeed, as shown with the
broad-band spectral energy distributions in Figure~\ref{fig:sed}, the
NIR/X-ray spectral index at $\approx 2.0$ d is $\beta_{\rm
NIR,X}\approx -0.70$ in agreement with our conclusion that the optical
and X-ray bands are located on the same synchrotron slope.  On the
other hand, the spectral index within the UV/optical/NIR bands,
$\beta_{\rm opt}\approx -0.85$, is steeper, indicating that
modest extinction is present.
The observed optical and X-ray decline rates require an electron
power law index of $p\approx 2.2$, in good agreement with the observed
spectral index. We note that in the Wind model $\nu_{\rm c}\propto t^{1/2}$
and hence once it crosses the X-ray band, the X-ray flux is expected
to decline as an unbroken power law, matching the observed behavior.

\begin{figure}
\begin{tabular}{cc}
\centering
 \includegraphics[width=0.48\columnwidth]{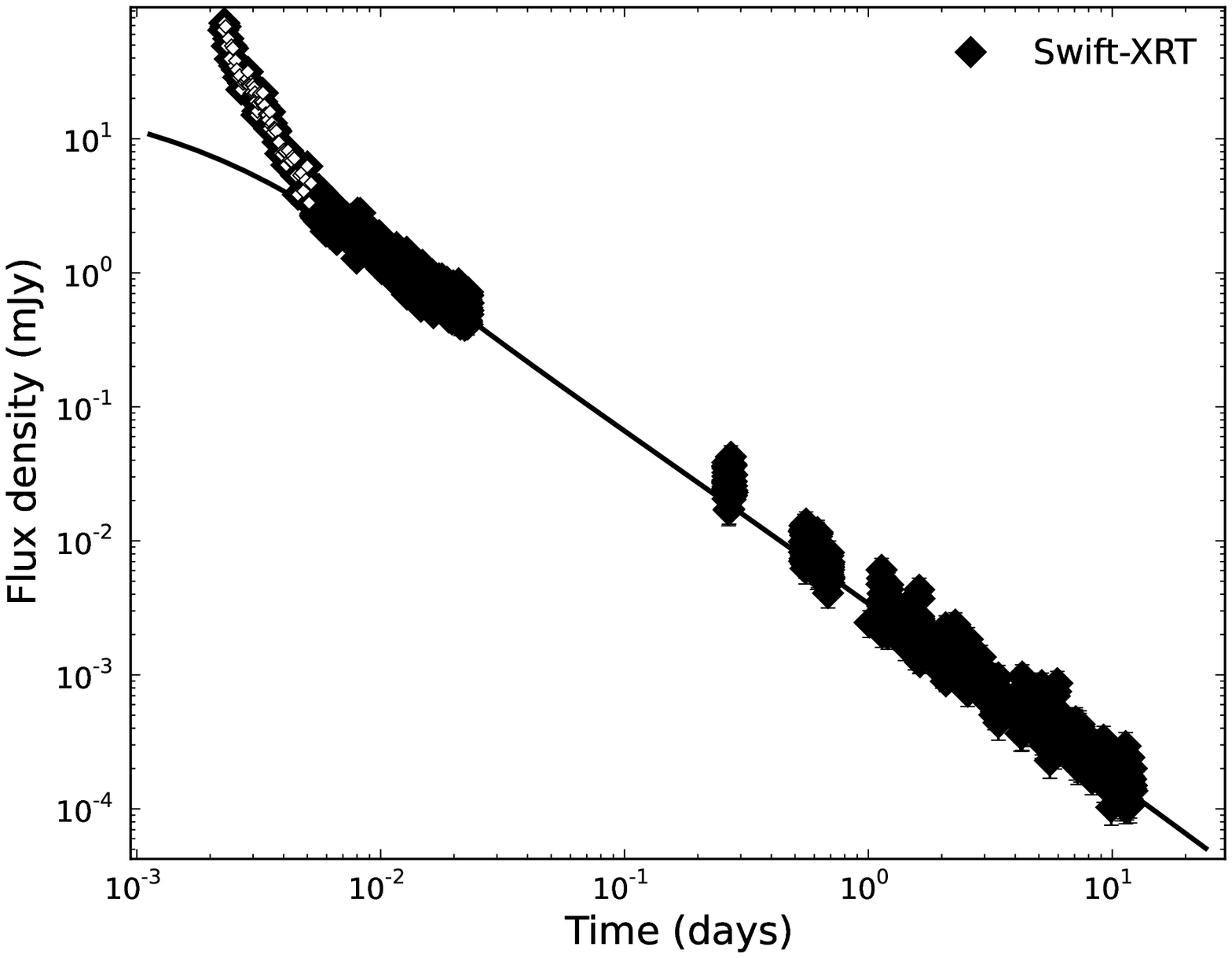}&
 \includegraphics[width=0.48\columnwidth]{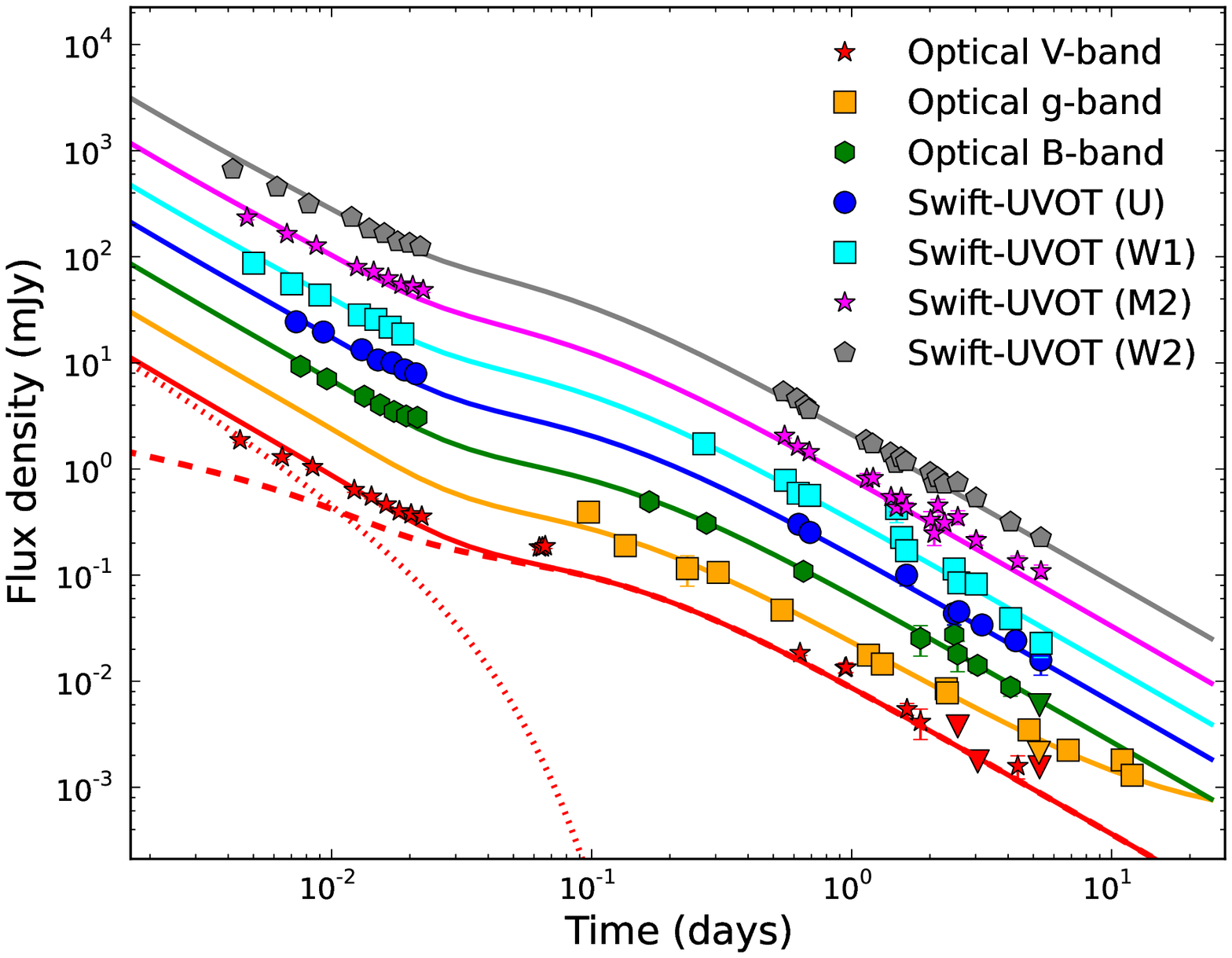}
\\
 \includegraphics[width=0.48\columnwidth]{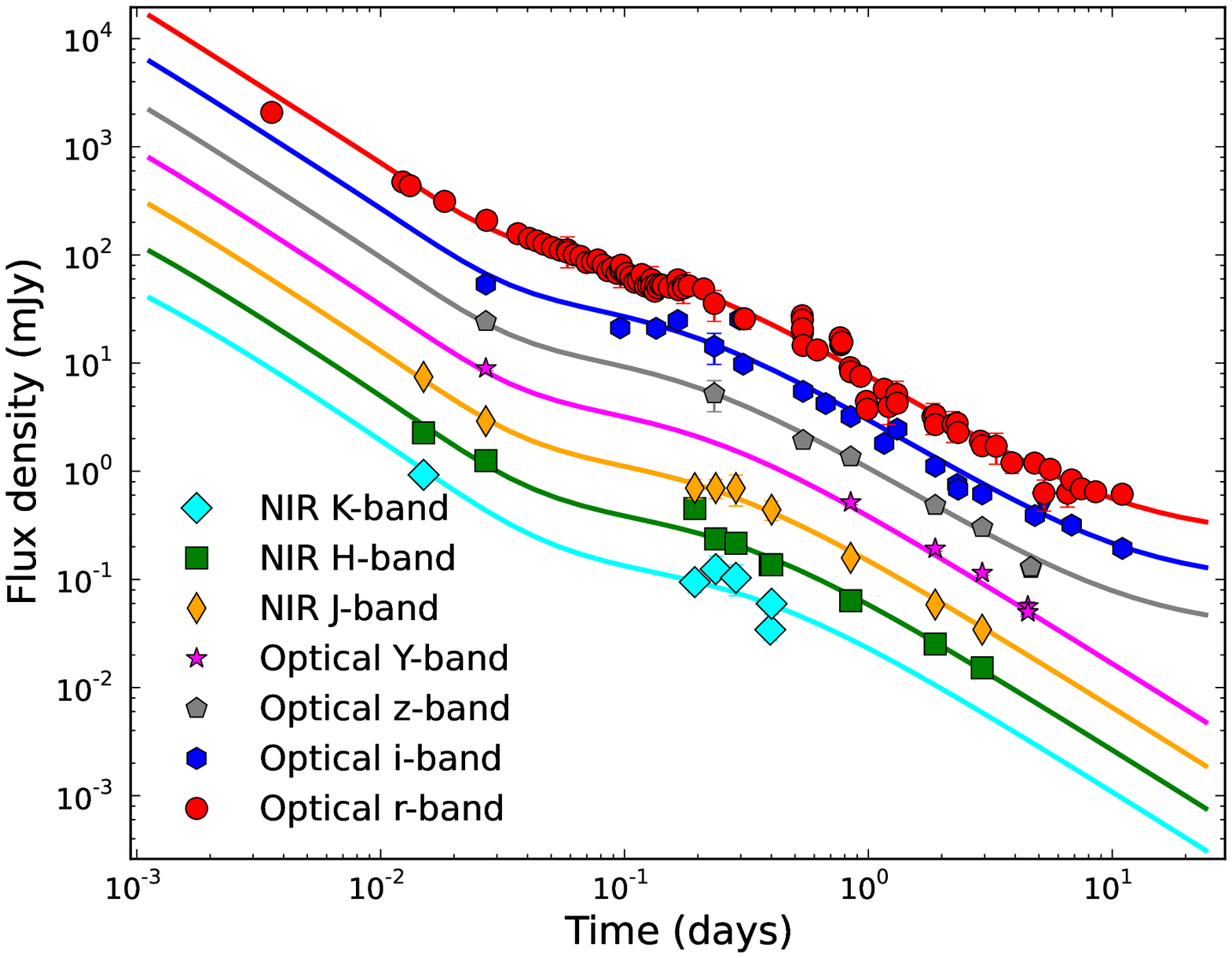} &
 \includegraphics[width=0.48\columnwidth]{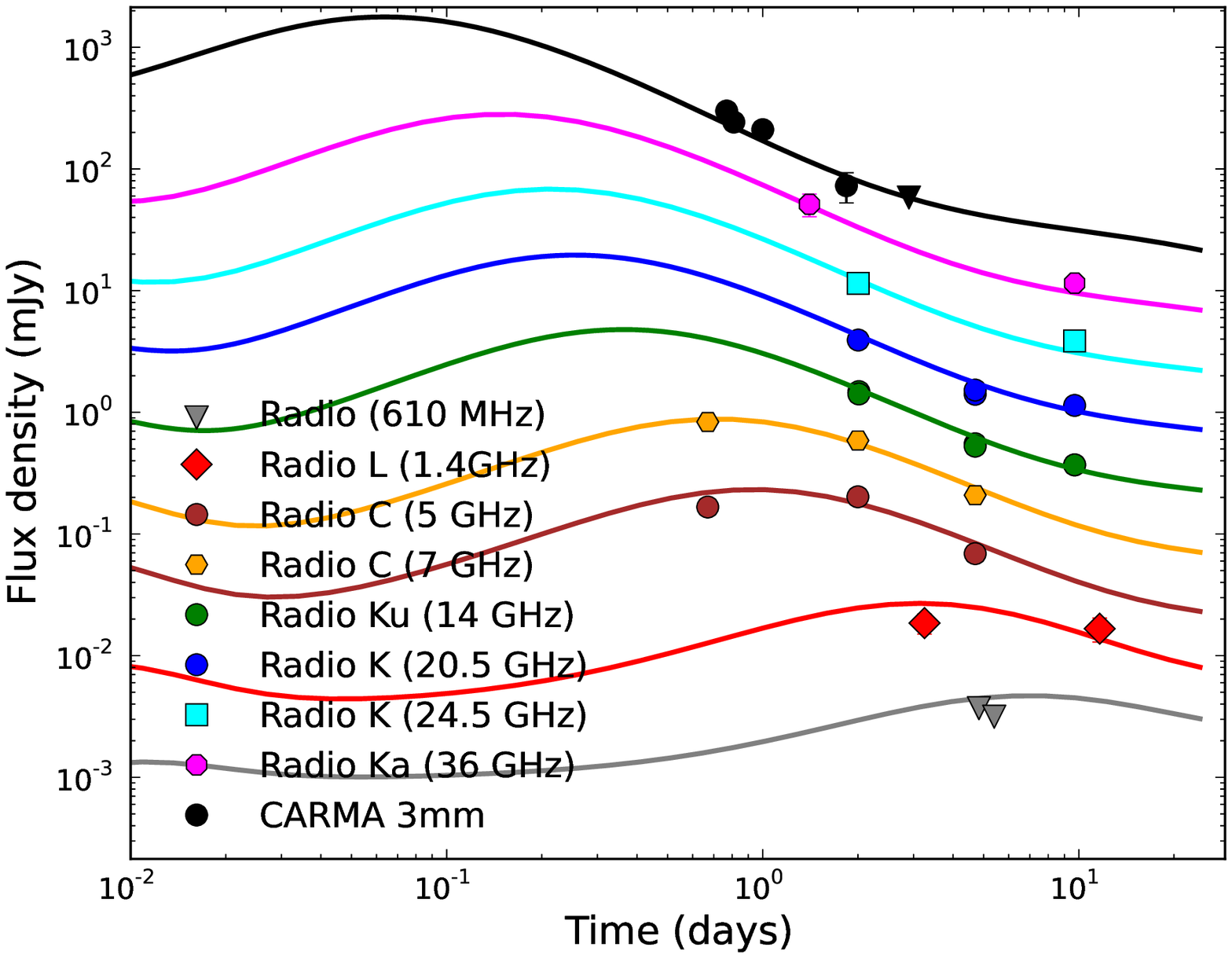}
\\
\end{tabular}
\caption{Light curves of the afterglow along with the combined reverse shock plus forward shock
model (lines). Top left: \Swift/XRT; top right: \Swift/UVOT and ground-based $g$-band; bottom left:
ground-based $rizYJHK$ observations from MMTcam and as reported in GCN circulars; bottom right:
GMRT, JVLA and CARMA observations spanning $0.6$ to $90$~GHz. Adjacent lightcurves have been offset
by a factor of 3 for clarity ($U$, $Y$, and the radio Ku band remain on the correct scale). In the
top right panel, we show a decomposition of the $V$-band light curve into reverse shock (dotted) and
forward shock (dashed) components to guide the reader. The \Swift/XRT data show a steep-to-shallow
transition around $\Delta t = 450$~s, which cannot be explained in our model. These data (open
symbols) are likely dominated by the low-energy tail of the prompt emission and we exclude them from
our fit. } \label{fig:lc}
 \end{figure}

Whereas the X-ray light curve follows a single power law decline, the
UV/optical/NIR light curves display a clear change in slope at
$\approx 0.1$ d (see Figure~\ref{fig:lc}).  The initial slope is
shallower, with $\alpha_{\rm opt}\approx -0.8$, and the time of the break is
chromatic, occurring later in the redder filters.  In the Wind model,
the passage of $\nu_m$ through an observing band results in a
transition from $\alpha=0$ to $\alpha=(1-3p)/4 \approx -1.4$ (for
$p=2.2$) in clear contrast to the observed evolution.  This indicates
that a different emission component dominates the UV/optical/NIR light
curves at $\lesssim 0.1$ d.

Even stronger evidence for a distinct emission component is provided
by the radio and millimeter data.  The relatively flat spectral index between the
radio/millimeter and optical bands at all times, $\beta_{\rm radio,opt}\approx -0.25$,
is inconsistent with a single power-law extrapolation from
the optical. This shallow slope cannot be caused by the location of
$\nu_m$ between the radio and optical bands because all light curves
below $\nu_m$ should be flat in the Wind model, while the observed
radio and millimeter light curves clearly decline at all frequencies
spanning $6.8$ to 90 GHz.
Similarly, the expected spectral index below $\nu_m$ is $\beta=1/3$ or
$2$ (the latter if $\nu_a$ is located above an observing band), while
we observe instead $\beta_{\rm radio}\approx -0.2$ at $\Delta t \approx 2$--5~d
(Figure~\ref{fig:sed}).

One possibility to flatten the spectral slope between the radio and
optical bands is to introduce a break in the electron energy
distribution at $\gamma_{\rm b}$ such that $N(\gamma)\propto
\gamma^{-p_{2}}$ (with $p_2<2$) for $\gamma_{\rm m}<\gamma<\gamma_{\rm
b}$.  For the observed value of $\beta_{\rm radio,opt}$ this requires
$p_2\approx 1.4$.  However, for this lower value of $p$ we expect a
shallower decline rate in the radio and millimeter of $\alpha\approx -0.8$, while
the observed decline in the millimeter bands is $\alpha_{\rm mm}\approx -1.5$.  This
argues against a break in the electron spectrum as the cause for the
shallow radio/optical spectral slope.

Instead, the distinct spectral and temporal behavior in the radio and millimeter
clearly requires a different synchrotron emission component, which is
also required to explain the UV/optical/NIR data at $\lesssim 0.1$ d.
We associate this component with the reverse shock, and show in detail
in the next section that it can explain the radio and millimeter data, as well as
the early optical data.  To gain insight into the reverse shock
spectrum we note that the SED at 0.67 d (Figure~\ref{fig:sed})
requires \numax\ to be located between $10$ and $90$ GHz, with
an optically-thin spectrum extending to the optical.  In addition, the
spectral index between 5.1 and 6.8 GHz, $\beta\approx 1.8$, is
indicative of self-absorption, with $\nu_{\rm a}\approx 7$ GHz.  The
optically thin spectrum observed in the radio at about 2.0 d indicates
that $\nu_a\lesssim 5$ GHz at that time.  Additionally, since the
emission from a reverse shock cuts off exponentially above $\nu_c$,
whereas we invoke some contribution from this component to the UV flux
at $\lesssim 0.1$ d, we require $\nu_{\rm c}\gtrsim 10^{15}$ Hz for the reverse
shock at $\approx 0.1$ d.

To summarize, the optical data at $\gtrsim 0.1$ d, when combined with
the full X-ray light curve at $\gtrsim 200$ s, require a Wind medium
with $p\approx 2.2$, $\nu_{\rm c}\sim\nu_{\rm X}$, and $\nu_{\rm m}\sim
\nu_{\rm opt}$ at $\sim 0.1$ d.  In addition, the radio and millimeter emission
and the early optical emission at $\lesssim 0.1$ d cannot be
accommodated with the same synchrotron spectrum, and therefore require
a separate emission component.  This emission component has $\nu_{\rm
a}\approx 7$ GHz and $\nu_m\approx 10-90$ GHz at 0.67 d, as well as
$\nu_{\rm c}\approx 10^{15}$ Hz at 0.1~d.  In the next section we
expand on these results with full broad-band modeling of the forward
and reverse shock emission.

\section{A Self-consistent Reverse Shock/Forward Shock Model}
\label{sec:rsfs}

Based on the basic considerations described in the previous section,
we construct a model SED for the radio to X-ray emission at 0.67 d
(Figure~\ref{fig:sed}).  The model is composed of two emission
components: (1) a forward shock (FS) which peaks between the
millimeter and optical bands, fits the NIR to X-ray SED, and provides
negligible contribution in the radio/millimeter; and (2) a reverse
shock (RS) which fits the radio to millimeter SED, and provides
negligible contribution at higher frequencies.  The synchrotron
parameters of the reverse shock are $\nu_{\rm a,RS}\approx 7$ GHz,
$\nu_{\rm m,RS}\approx 20$ GHz, $\nu_{\rm c,RS}\approx 10^{13}$ Hz,
and $F_{\nu \rm m,RS}\approx 10$ mJy.  The synchrotron
parameters describing the forward shock are $\nu_{\rm a,FS}<5$ GHz,
$\nu_{\rm m,FS}\approx 4\times 10^{13}$ Hz, $\nu_{\rm c,FS}\approx
2\times 10^{17}$ Hz, and $F_{\nu,{\rm m,FS}}\approx 3$ mJy.  Both SEDs are
in the slow cooling regime with the standard ordering of the
synchrotron break frequencies, $\nu_{\rm a}<\nu_{\rm m}<\nu_{\rm c}$.
We find that this combined RS plus FS model with a common value of
$p\approx 2.2$ completely describes the observed SED at 0.67 d
(Figure~\ref{fig:sed}).

We evolve both emission components in time to the other three epochs
where we have extracted multi-wavelength SEDs.
The evolution of the RS spectrum depends upon whether the shock is
relativistic or Newtonian in the frame of the unshocked ejecta.
\citet{zwd05} derive the temporal evolution of the RS break
frequencies, both before and after the ejecta crossing time (the
so-called deceleration time, $t_{\rm dec}$) in the two asymptotic
regimes of relativistic and Newtonian evolution.  An important
constraint on the shock evolution is provided by an $R$-band flux
density of about $77$ mJy measured by the Faulkes Telescope North at
$\approx 4.3$ min \citep{gcn14452}.  We find that a relativistic RS
over-predicts this early optical emission by about a factor of five.
In addition, in the relativistic case both $\nu_{\rm m,RS}$ and
$\nu_{\rm c,RS}$ evolve as $t^{-15/8}$, leading to a predicted decline
rate of $t^{-3(5p+1)/16}\sim t^{-2.25}$ above $\nu_{\rm \rm m,RS}$ (for
$p=2.2$), which is significantly steeper than the observed decline rate,
$\alpha_{\rm mm} \approx -1.5$. The Newtonian RS model, on the other
hand, offers an additional degree of freedom via the profile of the
shocked ejecta, $\gamma\propto r^{-g}$, where we expect $1/2<g<3/2$
from theoretical considerations \citep{zwd05}, although a value of
$g>3/2$ was found for GRB\,990123 \citep{sr03}.  We treat $g$ as a
free parameter and find a good match to the SED evolution with
$g\approx 5$. In the following, we focus on the Newtonian RS case.

Our implementation of the FS follows the smoothly-connected power-law
synchrotron spectrum for the Wind environment of \cite{gs02}, where we
compute the break frequencies and normalizations using the standard
microphysical parameters ($\epsilon_e$ and $\epsilon_B$), the
explosion energy ($E_{\rm K,iso}$), and the circumburst density in the Wind
profile ($A_*$).  We also use the SMC extinction curve \citep{pei92}
to model the extinction in the host galaxy ($A_{\rm V}$).  We use the
flux density of the host galaxy in the $griz$ filters as measured in
the Sloan Digital Sky Survey as an additive component to the relevant
filters.  Having determined a set of values for the break frequencies
of the RS spectrum at 0.67 d, we evolve these parameters based on the
Newtonian evolution, and fit separately for the FS contribution to the total
flux at all frequencies.  We determine the FS parameters using all of
the available photometry simultaneously with a combination of RS and
FS spectra.  To efficiently and rapidly sample the available parameter
space, we carry out a Markov Chain Monte Carlo (MCMC) analysis using a
python implementation of the ensemble MCMC sampler \emcee\
\citep{fhlg12}.

We find that the self-absorption frequency of the forward shock
($\nu_{\rm a,FS}$) declines below about $300$ MHz at 0.5 hr after
the burst and is not directly observed thereafter.  Consequently, the derived
blastwave parameters are degenerate with respect to $\nu_{\rm a,FS}$;
we quote all results in terms of this frequency, scaled to $14$ MHz
at 1~day (Table \ref{tab:fit}).  Imposing the theoretical
restrictions $\epsilon_e,\epsilon_B\le 1/3$ we further constrain the
blastwave parameters to the following narrow ranges: $12\,{\rm MHz}<\nu_{\rm
a,FS}<16\,{\rm MHz}$, $0.25<\epsilon_e<0.33$, $0.14<\epsilon_B< 0.33$,
$2.1\times 10^{-3}<A_*< 3.7\times 10^{-3}$, and $4.4< E_{\rm K,iso,52}<
5.8$.  Marginalizing the posterior density functions, we find the
following median values of the forward shock parameters: $p=2.23$,
$\epsilon_{\rm e}=0.30$, $\epsilon_B=0.20$, $A_*=2.9\times 10^{-3}$,
$E_{\rm K,iso,52}=4.9$, and $A_{\rm V}=0.18$ mag. The FS spectrum transitions
from fast cooling to slow cooling around $\Delta t \approx 1200$s.
The best-fit combined RS
and FS model is shown for the multi-epoch SEDs in Figure~\ref{fig:sed}
and for all available radio to X-ray light curves in
Figure~\ref{fig:lc}.  The model fully captures the observed evolution
across nine orders of magnitude in frequency and over three orders of magnitude in time.

The bulk Lorentz factor of the ejecta, $\Gamma$, can be calculated
from a knowledge of the deceleration time and the parameters of the
explosion.  In the internal-external shock model, we expect the
deceleration time to roughly match the duration of the burst.  Since
the \Swift/BAT $T_{\rm 90}\approx 163$ s and the optical flux
is already declining at $\approx 258$ s \citep{gcn14452} we take
$t_{\rm dec}\approx 200$ s.  Using the relation $t_{\rm dec}\approx
29(1+z)E_{\rm K,iso,52}\Gamma_{1.5}^{-1} A_{*}^{-1}$ \citep{zwd05}, we
find $\Gamma\approx 120-150$ at $t_{\rm dec}$,
where the range corresponds to the
uncertainty in $E_{\rm K,iso,52}$ and $A_*$ due to the uncertainty in
$\nu_{\rm a,FS}$.  For the median values of $E_{\rm K,iso,52}$ and $A_*$
reported above, we obtain $\Gamma\approx 130$.

In the preceding analysis, we independently determined the RS break
frequencies and fit for the parameters of the forward shock.  However,
the two synchrotron spectra are related since the two shocks propagate
in opposite directions from the contact discontinuity.  In particular,
we expect that at the deceleration time $\nu_{\rm c,RS}\sim\nu_{\rm
c,FS}$, $\nu_{\rm m,RS}\sim\nu_{\rm m,FS}/\Gamma^2$, and $F_{\nu,{\rm
m,RS}}\sim\Gamma F_{\nu,{\rm m,FS}}$.  For the parameters given above
we find that the FS parameters are $\nu_{\rm c,FS}\approx 8\times
10^{15}(t_{\rm dec}/200~{\rm s})^{1/2}$ Hz, $\nu_{\rm m,FS}\approx
2\times 10^{17}(t_{\rm dec}/200~{\rm s})^{-3/2}$ Hz, and
$F_{\nu,{\rm m,FS}}\approx 13(t_{\rm dec}/200~{\rm s})^{-1/2}$ mJy.  For the reverse shock, we have
$\nu_{\rm c,RS} \approx 2\times 10^{16}(t_{\rm dec}/200~{\rm
s})^{-1.3}$ Hz, $\nu_{\rm m,RS}\approx 5.3\times 10^{17}(t_{\rm
dec}/200~{\rm s})^{-1.3}(\Gamma/130)^{-2}$ Hz, and $F_{\nu,{\rm m,RS}}
\approx 10 (t_{\rm dec}/200~{\rm s})^{-0.9}(\Gamma/130)^2$ mJy, where
the power-law indices for $\nu_{\rm c,RS}$, $\nu_{\rm m,RS}$, and
$F_{\nu,{\rm m,RS}}$ are derived from the $g$-dependent expressions in
\citet{zwd05}.  We thus confirm that the expected relations between
the RS and FS parameters are satisfied to within a factor of two,
confirming our basic assumption that the two required emission
components indeed correspond to the reverse and forward shocks.

To summarize, a model that includes emission from the reverse and
forward shocks consistently explains all available data from radio to
X-rays and over a timescale of $\sim 200$ s to $\sim 10$ d.  The
resulting bulk Lorentz factor of the outflow at the deceleration time
is $\Gamma\approx 130$.

\section{Conclusions}
\label{sec:conc}

We present a detailed multi-wavelength study of the bright afterglow
of GRB~130427A spanning radio to X-rays.  From the optical and X-ray
data we conclude that the progenitor exploded into a Wind environment,
pointing to a massive star.  The radio and millimeter observations
present a spectrum and temporal evolution that cannot be explained by
emission from the forward shock alone.  We show that this emission is
consistent with synchrotron radiation from a Newtonian reverse shock.
With the available multi-band data this is by far the best example to
date of reverse shock emission in the radio/millimeter, particularly
when compared to previous detections that were based only on only 1--2
epochs at single frequencies \citep{kfs+99,bsfk03,sr03}.

Using multi-wavelength model fitting of the rich afterglow dataset,
with the well-sampled light curves spanning over three orders of
magnitude in time and nine orders of magnitude in frequency, we
determine the properties of the explosion and the circumburst medium.
In particular, we find a low circumburst density, with $A_{*}\approx
3\times 10^{-3}$, corresponding to a mass loss rate of
$\dot{M}=3\times 10^{-8}$ M$_{\odot}$ yr$^{-1}$ (for a wind velocity
of $1,000$ km s$^{-1}$).  The low density leads the reverse shock to
be in the slow cooling regime ($\nu_{\rm m,RS} < \nu_{\rm c,RS}$),
resulting in long-lived radio and millimeter emission.  We note that a
low density was also inferred for previous GRBs with likely radio
reverse shock emission (990123, 020405, and 021211; \citealt{clf04}),
suggesting that this is a requisite criterion for observable emission
from a reverse shock; in a high-density environment, the reverse shock
emission will decline rapidly due to efficient cooling of the
radiating electrons.

From the derived properties of the explosion and environment we obtain
the Lorentz factor of the outflow at the deceleration time,
$\Gamma(200\,{\rm s})\approx 130$, and show that the spectra of the
forward and reverse shock at the deceleration time are consistent with
theoretical expectations.  We infer $E_{\rm K,iso}\approx 7\times
10^{52}$ erg.  However, we note that the forward shock spectrum
transitions from fast to slow cooling at $\approx 1200$ s, and
radiative corrections between this time and $t_{\rm dec}\approx 200$ s
are a factor of a few \citep{sar97, dl98}.  This suggests that $E_{\rm
K,iso}\approx 2\times 10^{53}$ erg, which is $\approx 20\%$ of the
isotropic-equivalent $\gamma$-ray energy, implying that the radiative
efficiency was large.  The lack of an obvious break in the X-ray light
curves to $\approx 15$ d implies a lower limit of only $\theta_{\rm
jet}\gtrsim 2.5^{\circ}$ on the opening angle of the jet, and hence
$E_{\rm K,iso}\gtrsim 2\times10^{50}$ erg and $E_{\gamma,{\rm
iso}}\gtrsim 10^{51}$ erg.  Thus, due primarily to the combination of
a large isotropic energy and a low circumburst density, the lack of a jet
break at $\lesssim 15$ d is not surprising.

We conclude by noting that GRB\,130427A is likely to become the
benchmark for reverse shock studies in the JVLA and ALMA era.  Our
study demonstrates that a complete analysis of the explosion and
ejecta properties requires detailed multi-wavelength modeling to
leverage the anticipated exquisite data sets.
\acknowledgements

The Berger GRB group is supported by the National Science Foundation
under Grant AST-1107973.  CARMA observations were taken as part of the
CARMA Key Project, c0999, ``A Millimeter View of the Transient
Universe'' (PI: Zauderer) and VLA observations were taken as part of
programs 13A-046 (PI: Berger), 13A-41 (PI: Corsi), and SE0851 (PI:
Cenko).  The Karl G.~Jansky Very Large Array is operated by the
National Radio Astronomy Observatory, a facility of the National
Science Foundation operated under cooperative agreement by Associated
Universities, Inc. Support for CARMA construction was derived from the
states of California, Illinois, and Maryland, the James S. McDonnell
Foundation, the Gordon and Betty Moore Foundation, the Kenneth T. and
Eileen L. Norris Foundation, the University of Chicago, the Associates
of the California Institute of Technology, and the National Science
Foundation. Ongoing CARMA development and operations are supported by
the National Science Foundation under a cooperative agreement, and by
the CARMA partner universities.  GMRT is run by the National Centre
for Radio Astrophysics of the Tata Institute of Fundamental Research.
This work made use of data supplied by the UK Swift Science Data
Centre at the University of Leicester.

\bibliographystyle{apj}
\bibliography{/home/tanmoy/Projects/Edo/Papers/grb_alpha}

\begin{longtable}{l c c c c c}
\caption[GRB~130427A: Radio and millimeter observations]{\hbox {Radio and Millimeter Observations
of GRB~130427A}}\label{tab:photometry}\\
\hline \hline \\[-2ex]
   \multicolumn{1}{c}{\textbf{$t-t_0$}} &
   \multicolumn{1}{c}{\textbf{Observatory}} &
   \multicolumn{1}{c}{\textbf{Band}} &   
   \multicolumn{1}{c}{\textbf{Frequency}} &
   \multicolumn{1}{c}{\textbf{Flux density}} &
   \multicolumn{1}{c}{\textbf{Uncertainty}} \\      
   \multicolumn{1}{c}{(days)} &
   \multicolumn{1}{c}{} &
   \multicolumn{1}{c}{} &
   \multicolumn{1}{c}{(Hz)} &
   \multicolumn{1}{c}{(mJy)} &
   \multicolumn{1}{c}{(mJy)} \\[0.5ex] \hline
   \\[-1.8ex]
\endfirsthead

\multicolumn{6}{c}{{\tablename} \thetable{} -- Continued} \\[0.5ex]
  \hline \hline \\[-2ex]
   \multicolumn{1}{c}{\textbf{$t-t_0$}} &
   \multicolumn{1}{c}{\textbf{Observatory}} &
   \multicolumn{1}{c}{\textbf{Band}} &
   \multicolumn{1}{c}{\textbf{Frequency}} &
   \multicolumn{1}{c}{\textbf{Flux density}} &
   \multicolumn{1}{c}{\textbf{Uncertainty}} \\
   \multicolumn{1}{c}{(days)} &
   \multicolumn{1}{c}{} &
   \multicolumn{1}{c}{} &   
   \multicolumn{1}{c}{(Hz)} &
   \multicolumn{1}{c}{(mJy)} &
   \multicolumn{1}{c}{(mJy)} \\[0.5ex] \hline
   \\[-1.8ex]
\endhead

  \multicolumn{6}{l}{{Continued on Next Page\ldots}} \\
\endfoot

  \\[-1.8ex] \hline \hline
\endlastfoot
\num{0.67} & JVLA & C & \num{5.1e+09} & \num{1.5} & \num{0.075}  \\
\num{0.67} & JVLA & C & \num{6.8e+09} & \num{2.5} & \num{0.125}  \\
\num{2.00} & JVLA & C & \num{5.1e+09} & \num{1.82} & \num{0.091}  \\
\num{2.00} & JVLA & C & \num{6.8e+09} & \num{1.76} & \num{0.088}  \\
\num{2.00} & JVLA & K & \num{1.92e+10} & \num{1.31} & \num{0.0654}  \\
\num{2.00} & JVLA & K & \num{2.45e+10} & \num{1.28} & \num{0.0639}  \\
\num{2.00} & JVLA & Ku & \num{1.35e+10} & \num{1.48} & \num{0.0741}  \\
\num{2.00} & JVLA & Ku & \num{1.45e+10} & \num{1.42} & \num{0.0708}  \\
\num{4.70} & JVLA & C & \num{5.1e+09} & \num{0.621} & \num{0.0311}  \\
\num{4.70} & JVLA & C & \num{6.8e+09} & \num{0.626} & \num{0.0313}  \\
\num{4.70} & JVLA & Ku & \num{1.35e+10} & \num{0.552} & \num{0.0276}  \\
\num{4.70} & JVLA & Ku & \num{1.45e+10} & \num{0.527} & \num{0.0264}  \\
\num{4.70} & JVLA & K & \num{1.92e+10} & \num{0.469} & \num{0.0273}  \\
\num{4.70} & JVLA & K & \num{2.16e+10} & \num{0.508} & \num{0.029}  \\
\num{9.70} & JVLA & C & \num{7.29e+09} & \num{0.416} & \num{0.0352}  \\
\num{9.70} & JVLA & X & \num{8.4e+09} & \num{0.357} & \num{0.0434}  \\
\num{9.70} & JVLA & Ku & \num{1.35e+10} & \num{0.37} & \num{0.027}  \\
\num{9.70} & JVLA & Ku & \num{1.45e+10} & \num{0.37} & \num{0.021}  \\
\num{9.70} & JVLA & K & \num{1.92e+10} & \num{0.38} & \num{0.0475}  \\
\num{9.70} & JVLA & K & \num{2.45e+10} & \num{0.43} & \num{0.0376}  \\
\num{9.70} & JVLA & Ka & \num{3.6e+10} & \num{0.427} & \num{0.046}  \\
\midrule
\num{3.25} & GMRT & L & \num{1.39e+09} & \num{0.50}  & \num{0.10} \\
\num{4.83} & GMRT & 610 & \num{6.1e+08} & $<$ \num{0.30} & --- \\
\num{5.40} & GMRT & 610 & \num{6.1e+08} & $<$ \num{0.26} & --- \\
\num{11.6} & GMRT & L & \num{1.39e+09} & \num{0.45} & \num{0.1}  \\
\midrule
\num{0.77} & CARMA & 3mm & \num{9.25e+10} & \num{3.7} & \num{0.4}  \\
\num{0.81} & CARMA & 3mm & \num{8.5e+10} & \num{3} & \num{0.3}  \\
\num{1.00} & CARMA & 3mm & \num{9.25e+10} & \num{2.6} & \num{0.4}  \\
\num{1.84} & CARMA & 3mm & \num{8.5e+10} & \num{0.9} & \num{0.25}  \\
\num{2.90} & CARMA & 3mm & \num{8.5e+10} & $< $ \num{0.72} & --- \\
\end{longtable}

\begin{table}
  \centering\footnotesize
  \caption{Forward shock parameters for the best fit wind model}
  \label{tab:fit}\medskip
  \begin{threeparttable}
  \begin{tabular*}{0.5\columnwidth}[ht]{lc}
    \toprule
     Parameter & Value \\ 
    \midrule
    $p$           & 2.23\\
    $\epsilon_e$  & $0.30(\nu_{\rm a}/14~MHz)^{\nicefrac{5}{6}}$\\
    $\epsilon_b$  & $0.20(\nu_{\rm a}/14~MHz)^{\nicefrac{-5}{2}}$  \\
    $A_{*}$       & \num{3.0e-3}$(\nu_{\rm a}/14~MHz)^{\nicefrac{5}{3}}$\\
    $E_{\rm K,iso,52}$      & $4.9(\nu_{\rm a}/14~MHz)^{\nicefrac{-5}{6}}$ \\
    $A_V$         & 0.18~mag  \\
    $\Gamma$      & 130  \\
    $t_{\rm jet}$ & $\gtrsim 15$ days\\
    $\theta_{\rm jet}$ & $ > 2.5^{\circ}$\\
    \midrule
    $\nu_{\rm a,FS}$ & ($12$--$16$)$\times10^{6}$~Hz\\
    $\nu_{\rm m,FS}$ & $2.2\times10^{13}$~Hz\\
    $\nu_{\rm c,FS}$ & $2.8\times10^{17}$~Hz\\
    $\nu_{\rm a,RS}$ & $5.7\times10^{9}$~GHz\\
    $\nu_{\rm m,RS}$ & $1.2\times10^{10}$~Hz\\
    $\nu_{\rm c,RS}$ & $5.9\times10^{12}$~Hz\\
    \bottomrule
  \end{tabular*}
  \begin{tablenotes}[para,flushleft]
  All frequencies in this table are calculated at $\Delta t = 1$ day.
  The self-absorption frequency of the forward shock at 1 d,
  $\nu_{\rm a}$, is constrained to $12$~MHz $ < \nu_{\rm a,7} < 16$~MHz
   by the requirement \epse, \epsb $< \nicefrac{1}{3}$.
  \end{tablenotes}
  \end{threeparttable}
\end{table}

\end{document}